%% file: main.tex
\documentclass[journal]{IEEEtran}

\usepackage[utf8]{inputenc}
\usepackage{cite}
\usepackage{amsmath,amssymb,amsfonts}
\usepackage{graphicx}
\usepackage{booktabs}
\usepackage{array}
\usepackage{url}
\usepackage[hidelinks]{hyperref}

\hypersetup{
  pdftitle={When Agentic Trust Crosses Organizational Boundaries: Structural Externalization and a Reference Model for Trust Evidence},
  pdfauthor={Huafu Li and Jia Xia},
  pdfkeywords={agentic AI, trust evidence, cross-domain trust, provenance, authorization, evaluation}
}

\title{When Agentic Trust Crosses Organizational Boundaries: Structural Externalization and a Reference Model for Trust Evidence}

\author{Huafu~Li and Jia~Xia%
\thanks{Huafu Li and Jia Xia are with China Mobile Jiutian Artificial Intelligence Technology (Beijing) Co., Ltd. (e-mail: lihuafu@chinamobile.com).}}

\begin{document}
\maketitle

\input{sections/00-abstract}
\input{sections/01-introduction}
\input{sections/02-conceptual-method-and-related-work}
\input{sections/03-structural-externalization}
\input{sections/04-trust-control-requirements}
\input{sections/05-trust-evidence-envelope}
\input{sections/06-logical-reference-model}
\input{sections/07-analytical-scenarios-and-evaluation}
\input{sections/08-governance-limitations-agenda}
\input{sections/09-conclusion}

\bibliographystyle{IEEEtran}
\bibliography{ref}

\end{document}

%% file: sections/00-abstract.tex
\begin{abstract}
Agentic systems increasingly invoke tools, services, data, and other agents across organizational boundaries, yet a relying party cannot assess a delegated action solely from producing-domain controls and records. This paper develops Trustworthiness as a Service (TaaS) through a synthesis of trustworthy-AI governance, agent security, distributed trust management, identity, provenance, assurance, and control-plane research. The analytical unit is a cross-domain reliance proposition that names the issuer, subject and action, relying party, administrative boundary, evidence dependencies, adverse condition, and required verification or adjudication semantics. The three-condition structural-externalization diagnostic identifies propositions that depend on multiple domains, require producer-independent reliance, and must remain reviewable after revocation, failure, conflicting records, or dispute. For such propositions, the paper specifies a trust-evidence envelope: an immutable workflow manifest linked to append-only, issuer-attributed attestations for task-scoped authority, policy and execution decisions, provenance, validity, disclosure, status, challenge, and recovery. A topology-neutral logical reference model assigns these functions to explicit roles and trust domains. Three analytical scenarios and the TaaS-Eval protocol proposal define manifests, independent consumers, hard gates, adversarial evidence tests, metrics, and reproducible artifact reporting. By composing established identity, authorization, provenance, assurance, and governance mechanisms around a bounded delegated action, TaaS provides a reusable profile for cross-domain reliance. It makes evidence dependencies, independent verification, challenge, and recovery explicit, supporting interoperable governance and future evaluation without treating producer assertions as ground truth.
\end{abstract}

%% file: sections/01-introduction.tex
\section{Introduction}
\label{sec:introduction}

An enterprise agent receives authority to renew a cloud service and invokes a provider's provisioning API endpoint. The provider observes an authenticated request and an external effect on its resource. It does not observe the employee's original instruction, the limits placed on the agent, or whether an intervening tool or retrieved instruction changed the task. The enterprise can supply a runtime log, but that log is controlled by the same domain whose account the provider is being asked to accept. If the delegation is revoked after the request, the evidence conflicts with a provider-side record, or either status service becomes unavailable, the provider must still decide whether to honor, quarantine, reverse, or challenge the action. Its problem is evidentiary: what can it establish about this particular delegated action without treating the producer's assertion as its own proof?

Agentic workflows make this problem recurrent because they connect model outputs to tools, data, memory, other agents, and resources controlled by different organizations. Indirect prompt injection, poisoned retrieval, malicious tool descriptions, persistent memory compromise, and multi-agent communication attacks provide concrete paths by which an apparently authorized workflow can depart from its delegated task~\cite{debenedetti2024agentdojo,zhan2024injecagent,zou2025poisonedrag,wang2026mcptox,srivastava2025memorygraft,he2025redteamingmultiagent}. These studies identify failure modes rather than deployment prevalence. Risk frameworks and management-system standards, in turn, describe organizational functions for governing, assessing, monitoring, and responding to AI risk~\cite{nist2023airmf,nist2024genaiprofile,isoiec42001-2023,isoiec23894-2023}. Neither body of work by itself answers which evidence another organization should be able to verify for one bounded action at one administrative boundary.

Local control and cross-domain reliance are different requirements. Inside one administrative trust domain, an operator can authenticate its runtime, apply policy, sandbox tools, retain traces, and respond to incidents under one authority. Those controls remain necessary. A technical interface, process boundary, or API call does not alone create a need for external evidence semantics when the same authority controls the relevant trust roots, policy, primary records, and status functions. The requirement changes when a separately governed relying party must act on a proposition whose authority and evidence are divided across domains. The relying party then needs a basis for assessment that the producer cannot define and satisfy unilaterally, together with status or challenge semantics that remain meaningful after revocation, outage, conflicting records, or dispute.

Established mechanisms address important parts of this setting. Public-key infrastructure binds subjects to keys and supports certificate status; zero-trust architectures separate policy decisions from enforcement; distributed trust management and automated trust negotiation provide credential, delegation, and disclosure models~\cite{cooper2008rfc5280,rose2020zerotrust,abadi1992calculus,blaze1996decentralized,blaze1999keynote,winsborough2000automated}. Decentralized identifiers and verifiable credentials provide portable identifier and issuer-verifier semantics, while W3C PROV, in-toto, SLSA, SPDX, and C2PA provide reusable provenance or attestation structures~\cite{w3c2022didcore,w3c2025vcdm20,w3c2013provdm,torresarias2019intoto,slsa2025spec,spdx2024spec301,c2pa2025spec}. Governance, auditing, policy enforcement, metering, settlement, and incident response add further organizational functions. Trustworthiness as a Service (TaaS) integrates these antecedents rather than replacing them. The proposed profile supplies an agent-specific coordinate system that binds their outputs to a named subject, delegated task, action, relying party, trust boundary, and adverse condition.

In this paper, TaaS denotes an agentic trust-control profile comprising interoperable services and evidence schemas. Its central object is a trust-evidence envelope for a bounded delegated action. An immutable manifest identifies the workflow, parties, trust domains, and interpretation; independently attributable services append typed attestations for delegated authority, decisions, execution, provenance, assurance, status, disclosure, and challenge. The envelope is an interface contract, not a legal agreement, and a valid signature establishes neither event truth nor evidentiary completeness. A deployment that already exposes the required bindings and continuity semantics can satisfy the profile without adding a duplicate service or central platform.

Figure~\ref{fig:taas-overview} depicts the problem setting. A delegated action crosses an administrative trust boundary into a tool or resource domain. Delegation and authority attestations, execution evidence, and status or challenge paths must therefore form a coherent envelope that the relying domain can assess under its own accepted roots and policy.

\input{figures/fig-taas-overview}

The lower chain is deliberately not a single producer-issued verdict. Authority may be issued in the delegating domain, enforcement observed by the resource domain, and status or challenge records maintained by another accepted authority. The common envelope preserves those attributions and their causal references while allowing a verifier to reject an issuer, identify a missing dependency, or retain conflicting accounts. Independent verification therefore describes control over the acceptance procedure and its inputs; it does not promise an independent observation for every event.

The paper treats the problem as conceptual design, not system validation. A purposive, framework-driven synthesis supplies propositions from trustworthy-AI governance, agent security, identity and authorization, distributed trust management, provenance, assurance, and control-plane research. The analytical unit is a cross-domain reliance proposition: a claim about a subject or action that a named party must accept, reject, audit, price, challenge, or otherwise use across an administrative trust boundary. Each proposition is recorded with its issuer, subject and action, relying party, boundary, evidence and authority dependencies, adverse condition, and required verification or adjudication semantics. This unit prevents a broad label such as identity, provenance, or assurance from being classified as inherently external in every context.

The inquiry is organized by three research questions:

\begin{description}
\item[\textbf{RQ1.}] Under what conditions does an agentic trust proposition require evidence or verification semantics that no single administrative domain can define and validate unilaterally?
\item[\textbf{RQ2.}] What machine-readable evidence profile is needed to bind a delegated agent action to identity, task-scoped authority, execution decisions, provenance, validity, disclosure, and recovery across a named trust boundary?
\item[\textbf{RQ3.}] How can a relying party evaluate whether those outputs remain verifiable, usable, privacy-aware, and failure-responsive across trust domains?
\end{description}

The contribution reuses established identity, authorization, provenance, signing, policy-enforcement, audit, settlement, and incident-response primitives. Its scope excludes a prototype, benchmark result, theorem, exhaustive literature review, legal allocation of responsibility, and universal deployment topology. The analysis asks when existing mechanisms need shared cross-domain semantics around an agent action, specifies the evidence relations those semantics must preserve, and proposes how later implementations could be evaluated without using the producer's own account as the sole oracle.

The paper makes exactly three bounded contributions:

\begin{enumerate}
\item A \emph{structural-externalization diagnostic} that classifies a specific reliance proposition by multi-domain dependence, independent reliance, and continuity of verification or adjudication under an adverse condition. Its conclusion is limited to a need for interoperable reference semantics, not a central service or universal authority.
\item An \emph{agentic trust-evidence profile} that links an immutable workflow manifest to append-only, issuer-attributed attestations for delegated authority, policy and execution decisions, provenance, validity, disclosure, status, challenge, and recovery. The profile preserves disagreement and declared gaps without turning authentic producer claims into ground truth.
\item A \emph{logical reference model and TaaS-Eval evaluation protocol proposal}. The logical reference model separates roles and trust domains and admits centralized, federated, and protocol-mediated topologies. The protocol proposal pairs three analytical scenarios with tests of independent consumption, revocation, tampering, omission, replay, equivocation, disclosure, and recovery.
\end{enumerate}

Section~\ref{sec:conceptual-method-related-work} describes the conceptual method and antecedents; Sections~\ref{sec:structural-externalization}--\ref{sec:logical-reference-model} develop the diagnostic, profile, envelope, and logical reference model. Section~\ref{sec:analytical-scenarios-evaluation} specifies the scenarios and evaluation protocol, Section~\ref{sec:governance-limitations-agenda} examines governance and the research agenda, and Section~\ref{sec:conclusion} concludes.

%% file: figures/fig-taas-overview.tex
\begin{figure*}[t]
\centering
\includegraphics[width=\textwidth]{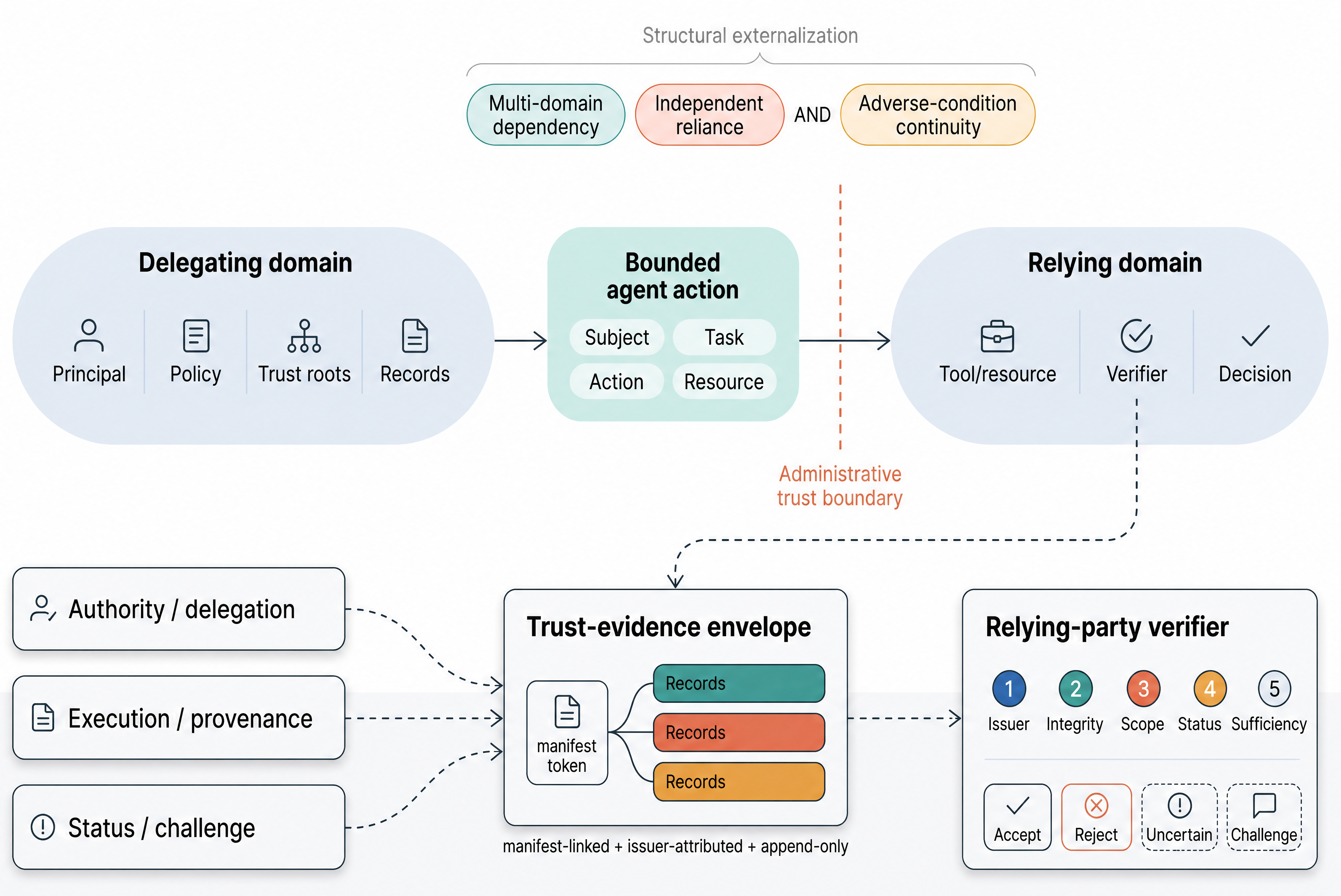}
\caption{For a proposition that satisfies the structural-externalization diagnostic, the trust-evidence envelope carries action-specific evidence across the administrative trust boundary for verification under the relying domain's accepted roots and policy.}
\label{fig:taas-overview}
\end{figure*}

%% file: sections/02-conceptual-method-and-related-work.tex
\section{Conceptual Method and Analytical Scope}
\label{sec:conceptual-method-related-work}

\subsection{Evidence scope and lineages}

This conceptual study uses a purposive, framework-driven synthesis of authoritative governance sources, reusable trust-control and evidence mechanisms, and agent-security studies relevant to cross-domain reliance. The corpus is non-exhaustive, and the cited attack studies establish plausible failure modes rather than their prevalence in deployed systems.

The synthesis draws on five evidence lineages. The first concerns trustworthy-AI properties and governance obligations. Ethics guidelines, risk-management frameworks, management-system standards, and auditing research describe duties related to reliability, privacy, accountability, oversight, and organizational control~\cite{jobin2019global,euhleg2019trustworthyai,nist2023airmf,isoiec42001-2023,raji2020accountability}. These sources identify what organizations should govern, but the present analysis asks how a relying party can evaluate a particular delegated action at a named boundary. The second lineage concerns agent threats. Studies of indirect prompt injection, poisoned retrieval, malicious tool descriptions, persistent memory compromise, and communication attacks document ways in which apparently authorized workflows can depart from a user's task or an operator's policy~\cite{debenedetti2024agentdojo,zhan2024injecagent,zou2025poisonedrag,wang2026mcptox,srivastava2025memorygraft,he2025redteamingmultiagent}. These studies are used to identify adverse conditions and evidence requirements, not to estimate prevalence.

The third lineage covers identity, authorization, and trust management. Public-key infrastructure, zero-trust architecture, distributed access-control calculi, KeyNote, and automated trust negotiation provide established models for subjects, credentials, policy compliance, delegation, and disclosure between parties~\cite{cooper2008rfc5280,rose2020zerotrust,abadi1992calculus,blaze1996decentralized,blaze1999keynote,winsborough2000automated}. Decentralized identifiers and verifiable credentials add portable identifiers and issuer-verifier credential exchange~\cite{w3c2022didcore,w3c2025vcdm20}. The fourth lineage concerns provenance and assurance. W3C PROV defines relations among entities, activities, and agents, while in-toto, SLSA, and C2PA specify attestations or provenance structures for software and digital content~\cite{w3c2013provdm,torresarias2019intoto,slsa2025spec,c2pa2025spec}. These mechanisms supply vocabulary and verification patterns without, by themselves, determining whether an agent's action was acceptable under a relying party's policy.

The fifth lineage comprises policy and control-plane architectures. Zero trust distinguishes policy decision and enforcement functions, while recent agent-oriented designs place runtime governance around tool calls, communication, and execution~\cite{rose2020zerotrust,kandasamy2025controlplane,tallam2026fiveplane,saga2026}. Adjacent TaaS and assurance formulations address compliance proof, network-service trust, or collaboration orchestration~\cite{scaramuzza2025zkmlops,scaramuzza2025zkmlopseng,li2025trustgain,zhu2026taas}. They reinforce the need for reusable trust evidence but do not fix the same unit of analysis adopted here.

Across these lineages, the present study treats the antecedents as composable mechanisms rather than deficient alternatives. PKI supplies subject and key validation; zero trust supplies policy-decision and enforcement separation; distributed trust management and trust negotiation supply credential, delegation, and disclosure models; DID and VC specifications supply portable issuer-verifier semantics; provenance and attestation systems supply causal and artifact-evidence structures; and AI governance and audit work supplies organizational control language. The agent-specific extension examined here is the binding of those outputs to a named subject, delegated task, action, relying party, trust boundary, and adverse condition.

\subsection{Unit of analysis and coding frame}

The unit of analysis is a \emph{cross-domain reliance proposition}. Such a proposition is a statement about an agent-related subject or action that a relying party must accept, reject, audit, price, challenge, or use when making a decision across an administrative trust boundary. Examples include whether an agent was authorized to invoke a provider's tool for a specified task, whether a disclosed trace belongs to the relevant execution, and whether a prior authorization remains valid after revocation. Focusing on propositions avoids assigning one fixed serviceability class to broad capabilities such as identity or provenance. An identity check consumed inside one organization and recognition of the same identity by another organization are distinct propositions with different evidence dependencies.

Each proposition is recorded through seven coding fields. Issuer identifies the party asserting or producing the relevant decision, credential, event, or evidence. Subject and action states who or what acted, on whose behalf, and what operation or resource is in question. Relying party identifies the actor expected to verify the proposition or act on it. Trust boundary names the point at which control or evidence crosses between administrative domains. The dependencies field records the evidence, policy, authority, keys, status services, and causal records needed to assess the proposition. The adverse-condition field identifies the case under which reliance is contested, such as revocation, outage, replay, conflicting records, producer compromise, or later dispute. The verification or adjudication semantics field states how the relying party checks status and meaning, and how disagreement can be resolved. These fields separate the authenticity of an assertion from its truth, policy acceptability, evidentiary sufficiency, and legal effect.

\subsection{Framework construction}

The conceptual model was constructed in four steps. First, the five lineages were read for propositions that require one party to rely on another party's identity, authority, decision, trace, status, or recovery claim. Each candidate was restated with the coding fields above, so the analysis concerned a specific issuer, action, relying party, and boundary rather than a capability name in the abstract.

Second, each proposition was subjected to the structural-externalization diagnostic developed in Section~\ref{sec:structural-externalization}. The diagnostic asks whether validity depends on more than one administrative domain, whether an outside party must rely independently of the producer, and whether verification, status checking, or adjudication must remain meaningful under an adverse condition. A positive classification calls for interoperable evidence or reference semantics. It does not prescribe a centralized service, ledger, or universal authority.

Third, the resulting design concerns were organized into the coverage framework labeled C1 through C12 and treated as a non-exhaustive checklist of trust-control concerns. The framework does not claim that the twelve labels partition all forms of trustworthy AI\@. Classification remains attached to a coded proposition and reliance context. The same capability can therefore contain local propositions and propositions that require cross-domain treatment.

Fourth, the logical reference model is applied to three worked scenarios: cross-organization tool invocation, a delegated multi-agent workflow, and post-incident evidence conflict and recovery. For each scenario, the analysis compares a local-control baseline with the proposed trust-evidence profile and records what an independent relying party can establish. The scenarios are analytical cases, not experiments. Their unresolved verification questions provide the requirements for the TaaS-Eval protocol proposal later in the paper.

\subsection{Validity boundary}

The method supports conceptual construction and internal consistency checking, not empirical effectiveness or field adoption. No independent coders, formal expert-validation panel, exhaustive corpus search, or inter-rater reliability study was used. The analysis makes no prevalence inference. The worked scenarios can expose ambiguity, missing semantics, and contradictory assumptions; they cannot substitute for implementation studies, adversarial trials, or evaluation by independent organizations. These limits bound claims for the diagnostic, evidence profile, reference model, and evaluation protocol.

%% file: sections/03-structural-externalization.tex
\section{Structural Externalization as an Operational Diagnostic}
\label{sec:structural-externalization}

\subsection{Roles and reliance semantics}

An \emph{administrative trust domain} comprises the principals, trust roots, policies, credentials, runtime controls, primary records, and status functions governed by one authority. The authority can change these elements without another domain's approval. A domain can coincide with an organization or with a separately governed unit inside one organization. A network hop, process boundary, or vendor API does not by itself create an administrative trust boundary.

The \emph{producer} originates an assertion or item of evidence. The \emph{issuer} is the identified authority that makes the assertion available under a stated verification method. One actor may hold both roles, but they remain analytically distinct when, for example, a runtime produces an event and an attestation service issues the corresponding record. The \emph{verifier} applies verification procedures to the assertion, supporting evidence, trust roots, and status information. The \emph{relying party} uses the resulting assessment to accept, reject, audit, price, challenge, or otherwise act on the target proposition. A relying party may perform verification directly or delegate it while retaining the decision.

The \emph{policy authority} controls the named rules under which a verifier evaluates issuers, schemas, authority, evidence, and status. The \emph{adjudicator} is a mutually recognized or legally competent person, institution, or procedure that resolves a contested record or interpretation when ordinary verification does not settle the disagreement. These roles can be held by the same organization, but combining them changes the degree of independence that a relying party can obtain.

Six semantic predicates must remain separate. Authenticity asks whether an assertion is attributable to its purported issuer under the stated verification method. Integrity asks whether the assertion or referenced evidence has changed since the relevant act of issuance or capture. Semantic conformance asks whether fields, types, units, and relations follow an agreed schema and interpretation. Integrity does not establish completeness, and semantic conformance does not establish that the described event occurred.

The remaining predicates concern use and consequence. Policy acceptability asks whether a verifier accepts the proposition under a named policy, audience, time, and operating context. Evidentiary sufficiency asks whether the available record supports a stated decision at the required level of assurance, including relevant absences or conflicts. Legal effect concerns the consequence given to the proposition by applicable law, contract, or recognized adjudication. Authenticity, integrity, or policy acceptance cannot determine that consequence on their own. Identity evidence leaves delegated authority undecided unless a delegation binds the subject to the action and scope. Provenance can describe an event sequence without determining responsibility, and a trace can inform adjudication without determining legal effect.

\subsection{Three-condition diagnostic}

The unit defined in Section~\ref{sec:conceptual-method-related-work} is a cross-domain reliance proposition, rather than a capability label. The structural-externalization diagnostic applies three conditions to that unit:

\begin{quote}
A reliance proposition requires cross-domain externalization when three conditions hold. First, its validity depends on evidence or authority controlled by more than one administrative domain. Second, at least one relying party must verify or act on it without granting the producer unilateral authority over both claim and evidence. Third, verification, status checking, or adjudication must remain available after revocation, failure, conflicting records, or dispute. These conditions require interoperable verification or adjudication semantics; they do not require a centralized service, public ledger, or universal third party.
\end{quote}

All three conditions are evaluated for the same proposition, relying party, and named boundary. The first condition is satisfied by a substantive dependency on separately controlled evidence or authority, not by data crossing a technical interface. The second asks whether the relying party has a basis for assessment that the producer cannot define and satisfy unilaterally. The third treats an adverse condition as a design test: the relevant status, records, verification rules, or dispute procedure must still be reachable and interpretable when ordinary operation has failed.

Application uses the seven coding fields established in the preceding section. The analyst identifies the issuer, subject and action, relying party, and administrative trust boundary. The analyst then inventories the evidence and authority dependencies by domain, selects a concrete adverse condition, and records the verification or adjudication semantics available under that condition. A positive classification requires all three conditions. A negative case clearly lacks at least one condition. A borderline case records a dependence that changes with the reliance purpose, accepted assurance basis, or institutional arrangement. Classification therefore attaches to the coded proposition and context, not to provenance, identity, metering, or any other capability in the abstract.

The diagnostic yields a limited conclusion. A positive classification identifies a need for interoperable semantics that an independent relying party can use. It does not show that an assertion is true, complete, acceptable under every policy, evidentially sufficient for another purpose, or legally effective. It also says nothing by itself about key protection, capture quality, service availability, privacy, governance competence, or implementation correctness. Conversely, a negative classification does not make service deployment unnecessary; it shows only that the stated proposition can be decided within one administrative domain. Table~\ref{tab:structural-externalization-cases} applies these distinctions to representative positive, negative, and borderline cases.

\input{tables/tab-structural-externalization-cases}

\subsection{Relabeling and evidence laundering}

The components used for cross-domain reliance have established lineages. Identity and access management, PKI, zero trust, trust negotiation, verifiable credentials, provenance, policy enforcement, auditing, metering, and dispute procedures already provide relevant objects and controls. Placing the same components behind service endpoints would amount to relabeling. The agent-specific contribution claimed here is narrower: the diagnostic identifies when a particular proposition about a delegated agent action needs externalization, and the proposed profile composes existing mechanisms around one subject, task-scoped authority, action, causal record, relying party, boundary, and adverse condition. An existing deployment that supplies these bindings and continuity semantics already meets the relevant profile; the model does not require a duplicate trust service.

Evidence laundering poses a separate objection. A well-formed signed envelope from a compromised runtime can launder false evidence. Its signature may support attribution and detect later alteration while the producer invents an event, omits a disconfirming record, or presents inconsistent accounts to different relying parties. Schema validity and a successful signature check therefore cannot serve as truth tests.

A design classified as positive must identify a basis beyond the producer's self-signature that is proportionate to the reliance decision. That basis can include evidence controlled by an independent source or counterparty, an effective challenge path with preserved conflicting records, or qualified assurance that states what was examined, by whom, under which assumptions, and with what limitations. The verifier must also consider source independence, collection conditions, omissions, status, and equivocation. These measures can increase justified confidence, but they still produce scoped assurance rather than certainty. Any legal consequence remains a matter for the applicable agreement, law, or adjudicator.

%% file: tables/tab-structural-externalization-cases.tex
\begin{table*}[t]
\caption{Representative applications of the structural-externalization diagnostic. Each classification is specific to the stated proposition, boundary, and adverse condition; it does not attach to the capability label.}
\label{tab:structural-externalization-cases}
\centering
\footnotesize
{\setlength{\tabcolsep}{3pt}
\renewcommand{\arraystretch}{1.14}
\begin{tabular}{@{}>{\raggedright\arraybackslash}p{0.18\linewidth}>{\raggedright\arraybackslash}p{0.27\linewidth}>{\raggedright\arraybackslash}p{0.22\linewidth}>{\raggedright\arraybackslash}p{0.27\linewidth}@{}}
\toprule
Case & Target proposition and boundary & Adverse condition & Classification and required semantics \\
\midrule
Cross-organization tool invocation & A provider decides whether an enterprise agent was authorized to invoke the provider's resource for the stated task at that time. The boundary is the delegating enterprise / tool provider boundary. & Revoked delegation or replayed request & \textbf{Positive.} Bind issuer, audience, task authority, resource, decision time, and request identifier; expose status, revocation, and challenge semantics. \\
Internal allow/deny decision & An enterprise uses its own policy decision point and enforcement point to decide whether an action was permitted by internal policy. No administrative trust boundary is crossed. & Policy update or runtime outage & \textbf{Negative.} Local policy version, decision, and logging semantics suffice for the stated proposition; a later external-certification proposition would be separate. \\
Selective provenance disclosure & An external auditor assesses whether a disclosed trace faithfully represents the execution relevant to the audit purpose. The boundary is the system operator / auditor boundary. & Selective omission, conflicting record, or producer compromise & \textbf{Borderline.} Classification depends on audit purpose and accepted assurance; define disclosure scope, lineage, status, challenge, and any independent corroboration. \\
Post-incident recovery proof & A counterparty assesses whether compromised authority was revoked, affected actions were isolated, and agreed recovery postconditions were met. The boundary is the affected operator / counterparty boundary, with a named adjudication forum. & Conflicting logs, unavailable status, or continuing dispute & \textbf{Positive.} Preserve timed revocation and recovery events, causal links, opposing evidence, verification status, and adjudication semantics. \\
\bottomrule
\end{tabular}
}
\end{table*}

%% file: sections/04-trust-control-requirements.tex
\section{Agentic Trust-Control Requirements and Coverage Framework}
\label{sec:trust-control-requirements}

The structural-externalization diagnostic in Section~\ref{sec:structural-externalization} applies to reliance propositions, not to service names. Its conditions yield requirements for an agentic trust-control profile. The C1--C12 vocabulary provides proposition-level coverage, while the lifecycle below arranges the corresponding decisions, evidence, and feedback.

\subsection{Semantic boundary and scope}

The semantic boundary used throughout the remainder of the paper is fixed as follows:

\begin{quote}
TaaS is an agentic trust-control profile comprising interoperable services and evidence schemas that bind a delegated agent action to a subject, task-scoped authority, execution decision, provenance, validity, disclosure, and failure or recovery semantics for a specified relying party at an explicitly named trust boundary.
\end{quote}

The word \emph{profile} is deliberate. TaaS specifies what independent producers and consumers must be able to represent and interpret; it does not prescribe a managed platform, a single control plane, or one deployment topology. It can be applied across application or runtime boundaries inside one organization, where a common authority controls the relevant trust roots, policy, primary records, and status functions. It can also span independent administrative domains. Only the latter case invokes structural externalization. A separately governed unit can constitute another administrative domain under the definition in Section~\ref{sec:structural-externalization}, but a process, runtime, or API boundary alone does not.

This profile is not a complete definition of AI trustworthiness. It addresses reliance on a delegated action and the evidence needed to review that action at a named boundary. Model capability, behavioral safety, fairness, human oversight, organizational competence, and legal compliance remain separate questions unless a specific proposition about them is issued and evaluated through the profile. Conformance to the profile therefore cannot establish that an agent, system, or organization is trustworthy in general.

\subsection{Requirements for a reliance decision}

The first requirement is an explicit reliance context. Every profile instance must name the relying party, the administrative trust boundary, and the purpose for which evidence will be used. It must also identify the policy authority and the accepted issuers, schemas, verification methods, and policy version that govern the decision. A generic audit audience is insufficient because acceptable evidence, disclosure, and failure behavior can differ between a tool provider, customer, auditor, and adjudicator.

Authority must be scoped to the action rather than inferred from identity alone. A delegation must identify its issuer, subject, audience, task, permitted action and resource, constraints, effective period, and revocation conditions. The resulting authorization decision must refer to that delegation and to the policy applied at decision time. Enforcement events must in turn refer to the authorization decision. These references allow a verifier to distinguish recognition of a subject from acceptance of that subject's authority for a particular task.

Assertions and events require attributable and connected evidence. Each assertion must identify its issuer and verification method. Runtime records must carry stable event identifiers, causal-parent references, and sufficient ordering information to relate an authorization request, decision, enforcement act, execution event, and observed effect. Independent verification means that a relying party can evaluate those records under its accepted trust roots and policy without allowing the producer to define both the claim and the acceptance test. It does not mean that every event needs a separate issuer. Where producer compromise is in scope, however, self-reported evidence needs counterparty evidence, an effective challenge path, or qualified assurance whose scope and limitations are stated.

Validity and disclosure remain part of the proposition after issuance. Credentials, policy decisions, attestations, and evidence references must expose issue and effective times, expiry where applicable, and status or revocation semantics. A relying party must be able to determine whether a record was valid for the event under review rather than merely valid when inspected. Selective disclosure must identify its audience, purpose, disclosure profile, transformation or redaction applied, and the capture scope or known gaps. This preserves a checkable relation to the underlying record without treating a hash as evidence of event truth or assuming that an incomplete view is sufficient for every purpose.

Each interaction also needs explicit failure and challenge semantics. The profile must state how a caller proceeds when evidence is absent, stale, replayed, contradicted, unavailable, or issued by a compromised service. It must distinguish refusal, degraded operation, quarantine, escalation, and later reconciliation rather than treating one response as universally correct. A challenge path identifies where a relying party can submit a conflicting record, request review, or obtain a remedy. If ordinary verification cannot settle the matter, the profile must name the agreement, authority, or procedure under which adjudication occurs. These requirements support review under adverse conditions without assigning legal effect to the evidence itself.

\subsection{Non-exhaustive proposition-level coverage}

The labels C1 through C12 are retained as a non-exhaustive vocabulary for checking whether a design has considered the trust-control concerns identified in the synthesis. They are neither a partition of AI trustworthiness nor a claim that every deployment needs twelve services. A single service may address several categories, and one category may require several issuers or enforcement points.

The first group covers recognition and authority. C1 identity trust asks whether a subject, organization, workload, or agent is recognized under accepted issuer, status, and assurance rules. C2 authorization control asks whether the subject has task-scoped authority for the named audience, action, resource, constraints, time, and revocation state. C3 intent governance adds the task and context assumptions used to decide whether the action remains within the delegated purpose. These categories are cross-domain only when another party must rely on the resulting recognition, authority, or task interpretation.

The second group covers execution and evidence. C4 reliable execution records the policy-versioned allow, deny, quarantine, degrade, or escalation decision and links it to an observable outcome. C5 behavioral provenance preserves event lineage, causal parents, ordering, capture scope, and known gaps, but it does not determine responsibility by itself. C6 observable assurance concerns scoped statements about control operation, method, period, assumptions, and limitations; an internal dashboard and an externally consumed assurance statement are different reliance propositions.

The third group covers disclosure, artifacts, and governance. C7 privacy and confidentiality requires an audience- and purpose-bound evidence view, with redaction, transformation, retention, and challenge semantics made explicit. C8 supply-chain integrity concerns model, tool, skill, data, and dependency identity under named artifact issuers and verification policy. C9 governance and compliance maps action evidence to obligations, control identifiers, scope, and review paths without turning the envelope into a clause-level compliance claim.

The final group covers consequence management. C10 metering and settlement separates local usage measurement from cross-party reconciliation of task-bound units, billing periods, anti-replay state, and counter-records. C11 incident response and recovery separates local containment from counterparty reliance on notification, revocation, affected-action boundaries, and recovery postconditions. C12 trust negotiation records the agreed issuer set, trust roots, schema and policy versions, disclosure terms, status endpoints, expiry, and revision history. In all twelve categories, the diagnostic attaches to the stated proposition, boundary, dependencies, and adverse condition rather than to the capability label.

\subsection{Trust-control lifecycle and feedback}

Figure~\ref{fig:trust-control-lifecycle} organizes the profile as a lifecycle rather than a dependency order. Governance and a trust agreement establish the accepted issuers, schemas, policies, and trust roots for a reliance context. Subject identity, task and intent, execution context, and delegation then become inputs to authorization. The authorization decision constrains enforcement, and enforcement governs the attempted action and its observable effects.

\input{figures/fig-trust-control-lifecycle}

Evidence is appended throughout these runtime stages. Authorization, enforcement, execution, tools, and observers can issue typed records without overwriting another issuer's assertion. Assurance, metering, and incident processing consume the resulting evidence under their own policies. Consumption does not imply acceptance: a verifier may reject an issuer, detect a gap, obtain a counter-record, or refer a disagreement to the stated challenge process.

Evidence review or assurance findings and incident findings can each initiate relying-party review; incident processing is not the sole trigger. The review has three distinct feedback paths. Revocation updates status and accepted authority for later authorization. Recovery reopens a relying party's acceptance only when the stated postconditions have been verified. Negotiation revises the agreement, accepted schemas, disclosure terms, status endpoints, policy versions, or trust roots when the parties' assumptions change. Each path can alter later decisions without rewriting the earlier evidence.

The coverage model and lifecycle are complementary. The coverage vocabulary identifies propositions and evidence semantics that require attention; the figure locates those propositions in the lifecycle and shows how later findings affect future decisions. Their mapping is many to many. In particular, identity is an authorization input rather than a universal trust anchor, and provenance is evidence for assurance or adjudication rather than proof of acceptable behavior. The lifecycle is also topology neutral: its roles may be implemented within one platform, federated across organizations, or connected through protocol-mediated services.

%% file: figures/fig-trust-control-lifecycle.tex
\begin{figure*}[t]
\centering
\includegraphics[width=\textwidth]{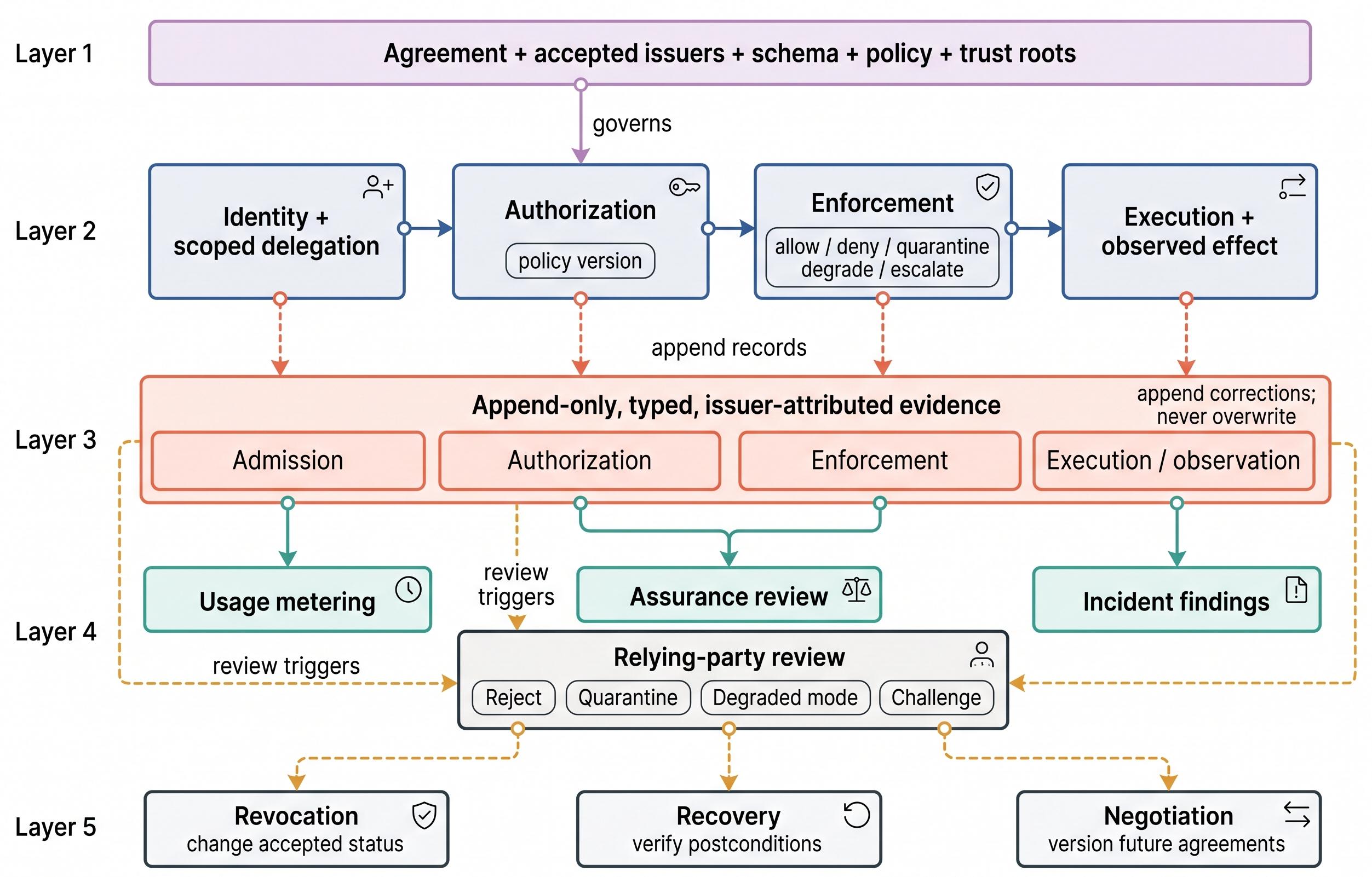}
\caption{The trust-control lifecycle separates runtime evidence consumers from three review-triggered feedback paths; arrows denote explicit control, evidence, review, or feedback relations, not capability sufficiency.}
\label{fig:trust-control-lifecycle}
\end{figure*}

%% file: sections/05-trust-evidence-envelope.tex
\section{Trust-Evidence Envelope}
\label{sec:trust-evidence-envelope}

\subsection{Definition and boundary}

The profile requires a stable object through which separately governed services can issue evidence about the same delegated action without collapsing their claims into one producer-controlled record. The trust-evidence envelope supplies that object:

\begin{quote}
A trust-evidence envelope is a machine-readable evidence and decision container for a bounded agentic action. It is an interface contract, not by itself a legal agreement. An immutable manifest identifies the workflow and trust domains; services append typed attestations and evidence references without overwriting another issuer's assertions.
\end{quote}

The term \emph{interface contract} refers to the syntax, interpretation, verification inputs, and failure behavior that producers and consumers agree to support. The envelope does not by itself create consent or an obligation, determine breach, allocate responsibility, select governing law, or provide a remedy. Legal effect arises only when applicable law, an external agreement, or a recognized adjudication procedure gives the record that effect. The envelope can preserve evidence for such a process without replacing it.

The design composes established elements and does not claim new identity, credential, provenance, or signing primitives. Issuer, verifier, audience, and status semantics draw on public-key infrastructure and verifiable credentials. Causal evidence relations draw on W3C PROV, while typed attestations draw on in-toto, SLSA, and C2PA~\cite{cooper2008rfc5280,w3c2025vcdm20,w3c2013provdm,torresarias2019intoto,slsa2025spec,c2pa2025spec}\@. The agent-specific profile binds these elements to one delegated action, its authority and policy decisions, the resulting events, a named relying party, and an explicit administrative trust boundary. A deployment that already provides those bindings can expose this profile without adding a duplicate control plane.

\subsection{Manifest and attestation semantics}

The manifest fixes the identity and interpretation of one envelope instance. It assigns stable \texttt{manifest\_id} and \texttt{manifest\_version} values and records its schema/version, issuer, verification method, integrity reference, workflow, and participating administrative trust domains. The manifest also fixes the subject, audience, relying party, reliance purpose, action/resource, creation time, disclosure profile, causal predecessor, and endpoint directory. The audience is authorized to receive a representation; the relying party uses the assessed proposition for the stated reliance purpose. These roles can differ.

Here, \emph{immutable} is a versioning rule rather than a prescribed storage technology. The pair \texttt{manifest\_id} and \texttt{manifest\_version} addresses one immutable version. A correction or renegotiation creates a new version with an explicit predecessor reference; a new workflow receives a new manifest identifier. Neither operation edits a manifest against which earlier attestations were issued.

Services add claims as append-only typed attestations. Every attestation has a stable \texttt{attestation\_id} and explicitly references the applicable \texttt{manifest\_id} and \texttt{manifest\_version}, in addition to the common identity, authority, causal, timing, disclosure, status, and challenge fields listed in Table~\ref{tab:trust-evidence-envelope}. A type-specific schema may state that a field is not applicable, but it cannot silently change the meaning of a common field. Stable attestation and event identifiers allow several issuers to describe the same action while preserving who asserted each fact.

Append-only typed attestations keep disclosed disagreements addressable. An authorization issuer cannot replace an execution record, and a runtime cannot rewrite an assessor's finding. Supersession, revocation, correction, counter-evidence, and recovery are later typed attestations that name the affected manifest version, attestation, event, credential, or evidence object. Targets cannot be inferred from append order. This rule is compatible with a database, federated repository, content-addressed object store, or transparency mechanism. It does not establish completeness or a consistent global history. A reliance context that includes omission or equivocation must add an independent witness, counterparty record, consistency mechanism, or adjudication procedure.

Evidence references keep potentially sensitive or large evidence objects outside the common envelope while making their relationship to an assertion explicit. A reference identifies the object, representation or schema, locator or retrieval method, integrity value when available, capture scope, and access or disclosure conditions. The evidence can remain under the originating domain's control, but the audience must be able to obtain the permitted representation and the verifier must be able to assess its availability and sufficiency for the named relying party and reliance purpose. An unavailable or intentionally withheld object remains a declared gap; its digest is not a substitute for access to relevant content.

\subsection{Proposed schema and verification}

Table~\ref{tab:trust-evidence-envelope} defines seven proposed envelope components. The manifest supplies the common coordinate system. Authority and policy attestations state why an action could proceed, while execution and provenance attestations state what an issuer observed or did. Assurance and status records qualify those claims over time. Disclosure views constrain what a particular audience receives, and challenge or recovery records preserve later disagreement and response.

\input{tables/tab-trust-evidence-envelope}

Verification is staged because a well-formed object can still be unacceptable. A verifier first checks the manifest issuer, verification method, and integrity reference, then resolves its schema and immutable identifier/version pair. For each attestation, it checks identifier uniqueness, the manifest reference, issuer trust, signature or digest integrity, and relevant time. It confirms that the audience may receive the disclosed representation and that the named relying party and reliance purpose match the intended use. The verifier then checks delegated scope, policy basis, causal references, evidence availability, disclosure gaps, expiry, and current or historical status before applying its acceptance and sufficiency policy. Passing an earlier stage does not imply acceptance at a later one.

Status must be evaluated for the event under review. The inspection-time status is also recorded, but it cannot substitute for historical state. A status response therefore names the affected manifest identifier/version, attestation, credential, event, or evidence object, together with status value, scope, effective time, issuer, and verification method. Revocation does not erase the original attestation, and it does not by itself decide whether a prior action was valid. The applicable policy must state the consequence of revocation, including how compromise dates, delayed publication, and later review affect historical reliance. If a required endpoint is unavailable, the envelope's failure semantics determine whether the verifier rejects, quarantines, uses a bounded degraded mode, or escalates.

\subsection{Disclosure, challenge, and evidentiary limits}

A disclosure view is an audience-specific, typed derivation linked to the source envelope. It records its purpose, included and omitted fields, redaction or transformation, evidence references, issue and expiry times, and the authority under which disclosure occurred. Audience limits who may receive the view; reliance purpose limits its permitted use, and the relying party makes the corresponding decision. The recorded linkage makes selective disclosure reviewable without implying that every relying party should receive the underlying trace. A verifier can reject over-disclosure as a privacy failure or treat under-disclosure as evidentially insufficient. Proof that disclosed fields correspond to a committed source still does not show that the source was complete.

A challenge record names the contested manifest identifier/version and each affected attestation, event, credential, or evidence identifier, together with the challenger, grounds, counter-evidence, requested remedy, and response deadline. A later resolution or recovery record targets the same identifiers and names the decision authority, outcome, verified postconditions, effective time, and any appeal or remedy path. Both remain alongside the contested assertion. The machine-readable challenge endpoint supports submission and status checking; the remedy path identifies the agreement, institution, or legal process competent to decide consequences that technical verification cannot determine.

Cryptographic checks have a narrower meaning than evidentiary acceptance. Under an accepted verification method and an uncompromised issuer key, a valid signature can support attribution and integrity of the signed representation. A hash can support integrity relative to an independently trusted digest and canonicalization procedure. Neither establishes truth, completeness, authorization, causal correctness, or absence of equivocation. These checks also depend on key binding, algorithm choice, key status at the relevant time, and protection of the verification inputs.

As discussed in Section~\ref{sec:structural-externalization}, a correctly signed envelope can still contain false or selectively incomplete attestations. The envelope exposes the identifiers, issuer, evidence, status, and challenge dependencies that a relying party must assess; it does not convert producer assertions into ground truth.

%% file: tables/tab-trust-evidence-envelope.tex
\begin{table*}[t]
\caption{Proposed trust-evidence envelope schema for a bounded agentic action and a named cross-domain reliance context.}
\label{tab:trust-evidence-envelope}
\centering
\footnotesize
{\setlength{\tabcolsep}{2.1pt}
\renewcommand{\arraystretch}{1.10}
\begin{tabular}{@{}>{\raggedright\arraybackslash}p{0.115\linewidth}>{\raggedright\arraybackslash}p{0.275\linewidth}>{\raggedright\arraybackslash}p{0.145\linewidth}>{\raggedright\arraybackslash}p{0.205\linewidth}>{\raggedright\arraybackslash}p{0.215\linewidth}@{}}
\toprule
Envelope component & Required fields & Issuer/owner & Verification rule & Failure or challenge semantics \\
\midrule
\multicolumn{5}{@{}>{\raggedright\arraybackslash}p{0.97\linewidth}@{}}{\emph{Common attestation fields:} stable \texttt{attestation\_id}; explicit \texttt{manifest\_id} and \texttt{manifest\_version} reference; schema/type/version; issuer and verification method; subject, audience, and action/resource; delegated authority and policy basis; decision or asserted event; evidence references and causal parent; issue/effective/expiry times; disclosure profile; status/revocation endpoint; challenge/remedy path.} \\[2pt]
Manifest & Stable \texttt{manifest\_id} and \texttt{manifest\_version}; schema/version; issuer and verification method; integrity reference; workflow; participants and administrative trust domains; subject; audience; relying party and reliance purpose; action/resource; creation time; causal predecessor; disclosure profile; endpoint directory & Workflow initiator or coordinating domain; ownership conveys no authority over later issuers & Verify issuer, method, integrity reference, and immutable identifier/version; match domains, audience, relying party, reliance purpose, action/resource, and predecessor & Unsupported version or unresolved parent causes rejection or quarantine; correction requires a new linked manifest version \\
Authority attestation & Common fields; delegation and credential references; action/resource scope and constraints & Delegator, credential authority, or other recognized authority & Validate issuer and delegation chain, manifest reference, audience, scope, time, constraints, and status under the accepted authority policy & Missing, stale, revoked, or out-of-scope authority causes denial or uncertainty; status and challenge records target its attestation ID \\
Policy/decision attestation & Common fields; policy authority, identifier, and version; request identifier; decision vocabulary and reason & Policy decision point operating for the named policy authority & Match request and decision; verify issuer, manifest reference, applicable policy version, evidence links, event time, and decision vocabulary & Policy mismatch or unavailable basis causes rejection or escalation; a counter-decision names the target attestation ID \\
Execution/\allowbreak{}provenance attestation & Common fields; event identifier/type; authorization and policy-decision references; outcome; ordering information; capture scope & Enforcement point, agent runtime, resource provider, tool, or independent observer & Check attribution, manifest reference, decision-to-event linkage, ordering, causal consistency, replay state, evidence integrity, and declared capture scope & Missing, replayed, contradictory, or unverifiable events are marked as gaps; counter-records target event and attestation IDs \\
Assurance/status attestation & Common fields; target manifest version, attestation, credential, event, or evidence identifier; status scope; assessed object/control; method, period, result, assumptions, and limits & Qualified assessor, status authority, or accepted assurance service & Verify target identifiers, issuer qualification, assessment scope and method, sampled evidence, time window, expiry, and historical status & Stale assurance, scope mismatch, revocation, or endpoint outage produces uncertainty rather than an inferred pass; unresolved cases escalate \\
Disclosure view & Common fields; source manifest version and attestation IDs; purpose; included/omitted field map; transformation or redaction; disclosure authority & Evidence custodian or authorized discloser & Check source identifiers, disclosure authority, audience, relying party, reliance purpose, declared omissions, transformation, and expiry & Over-disclosure is a privacy failure; under-disclosure can be insufficient; challenge targets the disclosure-view attestation ID \\
Challenge/\allowbreak{}recovery record & Common fields; contested manifest identifier/version and attestation, event, credential, or evidence IDs; challenger and decision authority; grounds and counter-evidence; remedy, status, outcome, postconditions, and appeal path & Challenger initiates; named adjudicator, incident coordinator, or recovery authority resolves & Verify every target identifier, standing and authority, causal links, counter-evidence access, procedural status, outcome, and recovery postconditions & Unresolved conflict or endpoint outage remains explicit; resolution targets the same identifiers and may escalate to the named agreement or legal process \\
\bottomrule
\end{tabular}
}
\end{table*}

%% file: sections/06-logical-reference-model.tex
\section{Logical Reference Model and Threat Assumptions}
\label{sec:logical-reference-model}

\subsection{Roles, trust domains, and topology}

The logical reference model assigns responsibility to roles rather than products. A \emph{principal or delegator} defines the task and grants authority to an \emph{agent runtime}. A \emph{resource provider} controls the tool, data, service, or physical resource on which the requested action operates. Each domain's \emph{policy authority} selects the rules for delegation, authorization, evidence acceptance, disclosure, and failure handling. An \emph{attestation issuer} attributes a typed assertion to itself under a stated verification method; the issuer need not be the runtime, enforcement point, resource, observer, or assessor that produced the underlying event.

A \emph{verifier} checks evidence under an accepted policy, while the \emph{relying party} decides whether and how to act on that assessment. An \emph{auditor or regulator} examines an authorized evidence view under its mandate. An \emph{incident coordinator} manages notification, containment, and recovery across affected domains. An \emph{adjudicator} resolves a contested record or consequence under a recognized agreement or procedure. These roles may be colocated when policy permits, but the allocation must remain explicit. In particular, combining producer, verifier, and relying-party authority changes the independence available for cross-domain reliance; combining incident coordination with adjudication can create a procedural conflict.

Figure~\ref{fig:taas-logical-reference-model} uses an illustrative allocation. Domain A is the delegating organization, Domain B is the resource organization, and optional Domain C supplies oversight or dispute functions. Domain C can be omitted when those functions are allocated elsewhere, but their endpoints and authority still need to be named when the reliance context requires them.

\input{figures/fig-taas-logical-reference-model}

The same relations admit centralized platform, federated, and protocol-mediated implementations. A platform can colocate roles while preserving issuer identities, tenant records, and policy scopes. A federation can leave policy, issuance, evidence, and status functions with their administrative owners. Protocol-mediated peers need no shared operator. A deployment-specific role directory maps each role to an authenticated endpoint and administrative owner. The mapping is versioned so a verifier can distinguish relocation or operator change from continuity of authority, and stale or ambiguous mappings invoke the negotiated failure policy. Established certificate, zero-trust, trust-negotiation, identifier, credential, provenance, and attestation mechanisms can implement parts of these relations~\cite{cooper2008rfc5280,rose2020zerotrust,winsborough2000automated,w3c2022didcore,w3c2025vcdm20,w3c2013provdm,torresarias2019intoto}\@. Evidentiary independence still depends on who controls the issuer, verification inputs, primary record, policy, and status function, not on topology alone.

\subsection{Pre-action agreement and trust administration}

Before an action, the participating domains name the relying party, reliance purpose, and administrative boundary. Issuer discovery and federation or verification endpoint discovery resolve accepted issuers, schemas, verification methods, credential and status services, evidence retrieval, and challenge or recovery contacts. The policy authorities negotiate or select policy versions, accepted trust roots, delegation semantics, disclosure terms, and behavior when an endpoint is unavailable. Discovery supplies inputs; the verifier still decides whether an endpoint and issuer are acceptable for the stated audience, action, and time.

Each domain owns its trust-root policy and appoints issuing authorities under that policy. The key lifecycle records issuance or enrollment, activation, permitted use, rotation, expiry, revocation, and destruction. Verification data retain the key identifier and validity interval needed to assess the event time. Rotation introduces a successor without rewriting earlier attestations, and revocation states an effective time whose consequence is set by relying-party policy. Root replacement requires authenticated distribution through a separately governed channel because a compromised root cannot repair itself.

Tenant and evidence isolation is a required property, not a consequence of assigning identifiers. Tenant identifiers and envelope namespaces scope policy and retrieval decisions; implementation controls must enforce separation of credentials, encryption contexts, storage, execution, status operations, and evidence access. An actor cannot read, append, revoke, or challenge another tenant's record without explicit authority. Selective disclosure further limits access by audience and reliance purpose. Section~\ref{sec:trust-evidence-envelope} defines the disclosure-view fields and declared-omission semantics.

\subsection{Runtime message and evidence flow}

The principal sends the runtime a delegation bound to the manifest version and intended action. Domain A's policy authority decides whether the runtime may issue the request. Domain B's policy authority evaluates the incoming authority and context, after which the resource provider enforces its own decision at the controlled boundary. The runtime and provider submit their respective decisions or observations to authorized issuers. Issuer A and Issuer B append separate attestations; neither can overwrite the other's assertion. Denial, quarantine, escalation, or degraded operation is recorded as an outcome rather than inferred from an absent success event.

Attestations follow the addressability, common-field, and explicit-expiry rules in Section~\ref{sec:trust-evidence-envelope}. The reference model adds message direction and ordering responsibility. Each issuer maintains unique event identifiers and attestation identifiers plus an issuer-local ordering value. Cross-issuer relations use request identifiers and causal-parent links from delegation to authorization, authorization to enforcement, and enforcement to observed effect. Receipt time does not create a total order; clock uncertainty, batching, and delayed delivery remain visible when they affect reliance.

\subsection{Post-action verification and response}

The attestation set and permitted disclosure view flow to the verifier, which also obtains evidence from any required independent source and queries the responsible status authority. The verifier applies the staged checks and historical-status rules in Section~\ref{sec:trust-evidence-envelope}, including explicit expiry and stale or revoked credentials, then sends its assessment to the relying party. A failed or uncertain check can lead the relying party to reject, quarantine, use a bounded degraded mode, or open a challenge. The incident coordinator manages cross-domain response; the adjudicator decides unresolved conflicts; and the recovery authority issues the resulting status or recovery record. Section~\ref{sec:trust-evidence-envelope} defines the target identifiers, counter-record, postcondition, appeal, and remedy fields. Earlier assertions remain available after resolution.

\subsection{Threat assumptions and residual guarantees}

A compromised issuer can sign a false statement, and colluding issuers can present consistent but false records. Replay reuses a valid record outside its request, audience, or validity context; equivocation presents incompatible histories to different parties; omission withholds a relevant record. Identifiers, nonces, expiry, audience binding, issuer-local order, causal links, and consistency comparison can expose some instances, but the envelope cannot establish completeness or truth. Where these threats affect the reliance decision, policy must require counterparty or independent evidence, scoped assurance, or adjudication. Those measures do not defeat collusion among all accepted sources.

Privacy leakage remains possible after successful verification. Cross-tenant access, excessive disclosure, stable correlators, or challenge material can reveal sensitive identity, task, or provenance data. Enforced tenant isolation, purpose-bound access, data minimization, selective disclosure, and retention limits constrain exposure, but their effectiveness depends on implementation and governance.

Service outage has policy-specific consequences. The pre-action agreement states which unavailable issuer, verification, status, evidence, challenge, or adjudication endpoint causes refusal or quarantine and which permits a time-bounded degraded mode. A degraded decision records the missing dependency, cache age, expiry bound, and reconciliation requirement. Recovery does not erase actions taken during the outage, and an unavailable dispute function leaves the matter explicitly unresolved.

Governance capture and trust-root compromise exceed what record syntax can prevent. A captured policy authority can authorize harmful conduct under a valid policy version; a compromised root can make unauthorized issuers appear valid. Separation of ownership, issuance, verification, audit, incident coordination, and adjudication may limit concentration only when those roles are independently governed. Root compromise requires out-of-band replacement, distrust or revocation of affected descendants, preservation of the suspected interval, and reassessment of decisions made during it. If the policy owner, root owner, issuers, verifier, relying party, and adjudicator collude, the model can expose their stated roles and records but cannot supply an external basis for truth or remedy.

The model therefore specifies responsibilities, message direction, and observable failure states. It does not claim to prevent compromise, collusion, privacy leakage, outage, governance capture, or root compromise.

%% file: figures/fig-taas-logical-reference-model.tex
\begin{figure*}[t]
\centering
\includegraphics[width=\textwidth]{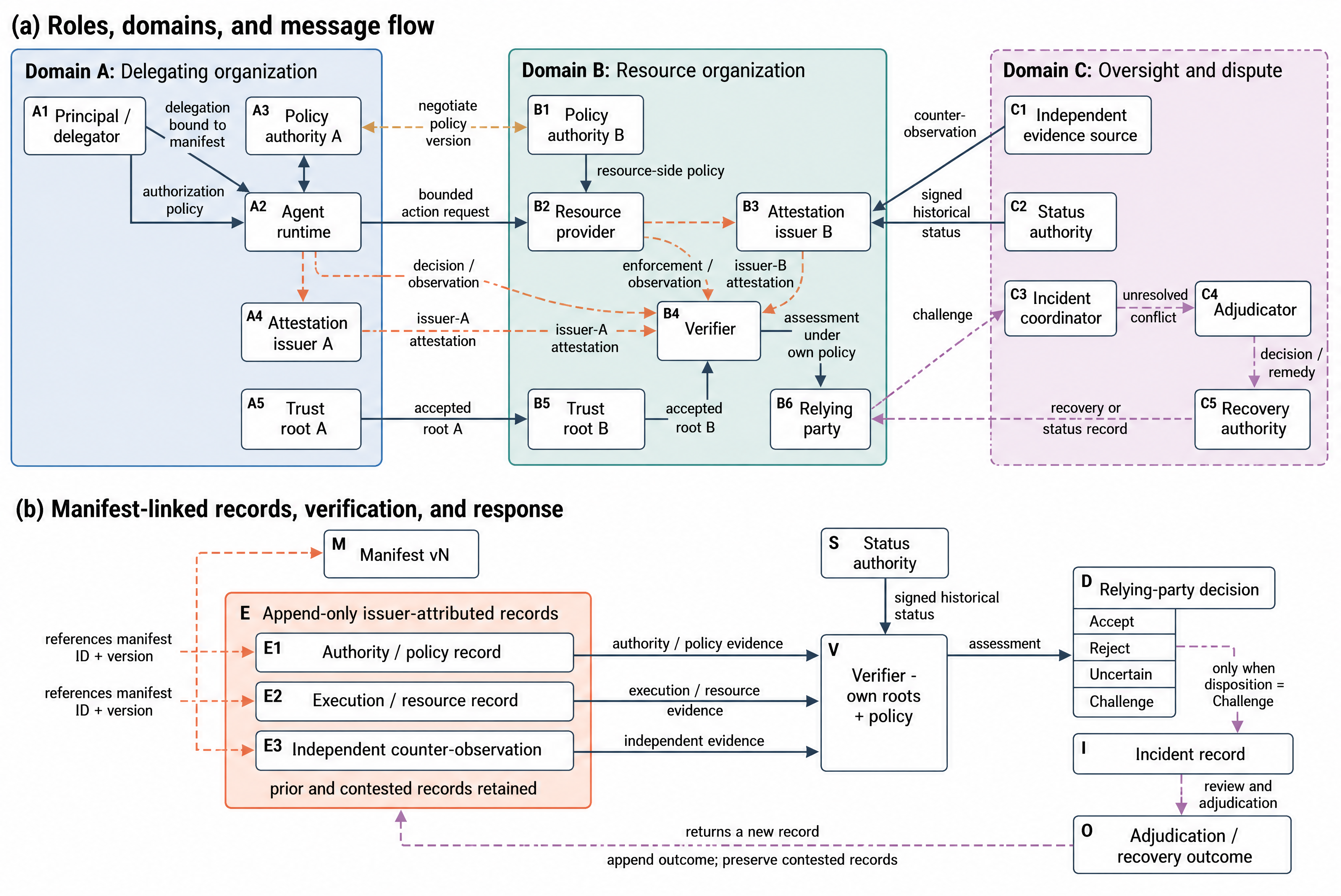}
\caption{Topology-neutral logical reference model: panel (a) separates required operational roles from optional oversight roles; panel (b) preserves manifest-referenced, issuer-attributed records through verification and response.}
\label{fig:taas-logical-reference-model}
\end{figure*}

%% file: sections/07-analytical-scenarios-and-evaluation.tex
\section{Analytical Scenarios and TaaS-Eval Protocol}
\label{sec:analytical-scenarios-evaluation}

\subsection{Analytical scenarios}

Three analytical scenarios apply the structural-externalization diagnostic to specific reliance propositions. Local controls remain necessary in every scenario; the question is whether a separately governed relying party can assess the stated proposition without accepting the producer's account as its own proof.

\subsubsection{Cross-organization tool invocation}

An agent runtime operated by a customer organization produces a request to a tool controlled by a service provider. The provider-side verifier checks the request for the resource owner, which is the relying party. The customer's delegation authority and the provider's access-policy authority govern different parts of the decision; the administrative boundary lies between those organizations. A local baseline can authenticate the runtime, restrict its tools, and retain a customer-side execution log. Those controls cannot establish for the provider that the requested action was delegated for this task, remains within scope, or was evaluated under the policy versions named in the request.

The cross-domain reliance proposition is that the identified runtime invoked the resource under valid task-scoped authority and that the provider enforced its recorded decision. The envelope therefore needs a manifest, an authority attestation, customer- and provider-side policy decisions, a provider execution attestation, causal evidence references, applicable status records, and a challenge path. If the customer runtime is both producer and attestation issuer, compromise lets it sign a false description of user intent or omit an earlier instruction. The provider can still verify attribution and compare the self-report with delegation records and its own enforcement observation, but it cannot establish the user's actual intent from the runtime signature. Missing independent evidence leaves that part uncertain; contractual consequence or a contested interpretation remains for the named adjudication procedure.

\subsubsection{Delegated multi-agent workflow}

In the second case, an orchestrating agent in one organization produces a subtask delegation for an agent operated by another organization, which may call a third organization's tool. The delegating organization's verifier assesses the returned envelope for the workflow owner. Policy authority is divided among the original delegator, the downstream agent operator, and any resource provider, with an administrative boundary at each organizational handoff. The local baseline records the orchestrator's plan and applies access control to its own agent. It does not expose how the downstream operator interpreted the subtask, attenuated authority, or controlled a later tool call.

The cross-domain reliance proposition is that every downstream action remained within the transitive delegation and that the returned result is causally linked to those authorized actions. Required records include the workflow manifest, one authority attestation for each delegation edge, policy and runtime decisions at each governed boundary, execution or resource attestations, causal-parent links, status, disclosure, and challenge records. A compromised downstream operator that issues its own completion attestation can conceal an undeclared tool or fabricated result behind a well-formed self-report. Counterparty receipts or independent resource observations can expose some discrepancies, but signatures and an internally consistent trace do not establish semantic faithfulness when accepted sources collude. Conflicting delegation interpretations, omitted branches, and allocation of responsibility therefore remain matters for challenge and, when technical checks cannot resolve them, adjudication.

\subsubsection{Post-incident evidence conflict and recovery}

After an incident, a customer runtime may produce an attestation that an action was blocked while the provider's enforcement point produces a record of an external effect. A jointly accepted verifier examines both accounts for the two relying organizations. The customer and provider policy authorities use the versions named in their pre-incident trust agreement, which also identifies status services, an incident coordinator, a recovery authority, and an adjudicator. The boundary separates the customer and provider domains, with an optional oversight domain for independent evidence and dispute handling. Each organization can use its local logs for containment, but neither local account settles the cross-party proposition when the records conflict.

The cross-domain reliance proposition now concerns which action occurred, which authority and decision governed it, when revocation became effective, and whether the agreed recovery postconditions were met. The evidence set includes the original manifest and authority, decision, execution, status, revocation, counter-record, challenge, and recovery attestations. A compromised runtime may issue its own containment claim, or a self-reporting recovery issuer may attest to cleanup without evidence from the affected resource. The verifier can test identifiers, timing, causal consistency, status, and the stated postconditions against available counter-records. It must leave an unavailable record or unresolved contradiction explicit. Event truth under collusion, legal fault, and the sufficiency of a remedy exceed the envelope and require the adjudicator or other competent process named in the agreement.

\subsection{TaaS-Eval: an evaluation protocol proposal}

TaaS-Eval specifies how future implementations can evaluate the trust-evidence envelope associated with a bounded agentic action. Existing tool-use and harmful-action benchmarks can supply task and threat seeds, but a seed enters this protocol only after its trust boundary, relying proposition, required evidence, and oracle have been specified~\cite{debenedetti2024agentdojo,zhan2024injecagent,andriushchenko2025agentharm}\@.

\subsubsection{Manifest, execution, and consumers}

Each evaluation begins with a versioned task and threat manifest. The manifest fixes the task cases, valid and invalid inputs, subject and action vocabulary, administrative trust boundary, and system under test (SUT). It also fixes the profile and policy versions, accepted issuers and trust roots, required envelope attestations, adversarial injections, and expected failure modes. The manifest identifies the oracle and the evidence available to that oracle. The SUT is recorded as a service or an implementation of the agentic trust-control profile, together with its model, agent runtime, tool environment, policy configuration, and component versions. Configuration digests support comparison of runs but do not establish that the configuration behaved as declared.

The oracle is fixed before execution and cannot be derived solely from the SUT's output. Controlled tasks can use a known authorization matrix, instrumented resource effects, deliberately constructed valid or invalid envelopes, a disclosure allowlist, and specified recovery postconditions. A status oracle is owned by the harness or issued by an authority outside the SUT, even when the SUT also emits status records. For a contested field without independent ground truth, the oracle records the field as unresolved rather than assigning a true or false ground-truth label. The task manifest predeclares the expected terminal reliance disposition for that unresolved case. This distinction prevents an authentic producer self-report from becoming the oracle for its own claim.

Two validation predicates are recorded before the consumer assigns a terminal reliance disposition. Schema-valid means that an item parses under the declared schema and satisfies its field, type, identifier, and reference constraints. Cryptographically authenticated means that the representation and purported issuer pass the declared cryptographic verification step. Neither predicate establishes current authority, status, truth, or sufficiency. After the authority, status, causal, evidence, and disclosure checks, the independent consumer assigns exactly one terminal reliance disposition for the named audience and purpose: verified-for-reliance, rejected, uncertain, or escalated. Verified-for-reliance applies only when the consumer accepts the proposition under its own policy and trust inputs. Escalated ends the current consumption decision and transfers the case to the predeclared challenge or adjudication procedure. An item may satisfy both validation predicates and still receive any of the latter three dispositions.

Stochastic agents follow a predeclared repeated-run policy. Before any output is inspected, the policy fixes the number of repetitions, seeds where controllable, sampling settings, and retry and timeout rules. It also fixes environment reset and the aggregation rule. Every repetition retains its task instance, randomization inputs, envelope, and external observations. Its execution outcome and terminal reliance disposition are recorded separately. The adversarial harness injects conditions at their declared point before or during execution and envelope capture. The trace validator then applies deterministic schema, identifier, authority, causal-link, timing, disclosure, and status checks. An independent consumer evaluates the permitted envelope view with separately administered trust roots and policy inputs and without access to private runtime state. Figure~\ref{fig:taas-eval-protocol} gives the resulting sequence.

\input{figures/fig-taas-eval-protocol}

Four hard gates precede score reporting. The identity and authority gate requires a verifiable subject, task-scoped delegation, audience, action and resource scope, validity, and status. The runtime-decision gate requires a policy-versioned decision linked to an observable enforcement outcome, including denial, quarantine, degraded operation, or escalation. The independent-consumption gate requires the consumer to return the oracle-expected terminal reliance disposition for every required item without private SUT state. The failure-handling gate requires behavior consistent with the manifest for invalid, stale, conflicting, privacy-violating, or unavailable evidence, together with challenge and recovery records when applicable. A task cannot receive an overall pass if any hard gate fails.

Table~\ref{tab:taas-eval-protocol} maps these gates to concrete checkpoints. A gate vector $\mathbf{g}=(g_1,g_2,g_3,g_4)$ records their Boolean outcomes. Scores remain a vector rather than being collapsed into a universal scalar. If a study uses weights or thresholds, its manifest must declare them and its missing-data rules before execution; aggregation cannot override a failed gate.

\input{tables/tab-taas-eval-protocol}

\subsubsection{Metrics and artifact reporting}

Metric denominators are part of the task manifest. Every ratio defined here and in Table~\ref{tab:taas-eval-protocol} is reported with its raw numerator and denominator. A zero denominator yields N/A, not zero. Let $A_{\mathrm{req}}$ count oracle-valid required attestation opportunities, including a required item that the SUT omits, and let $A_{\mathrm{iv}}$ count the subset for which the independent consumer reaches verified-for-reliance. Independent-verification success is $\mathrm{IVS}=A_{\mathrm{iv}}/A_{\mathrm{req}}$. An oracle-invalid compromised-producer attestation can be schema-valid and authenticated, but it is excluded from $A_{\mathrm{iv}}$ and cannot count as verified-for-reliance.

Execution and reliance errors use different units. Execution false acceptance is
$\mathrm{FAR}_{\mathrm{exec}}=$ oracle-invalid action requests that produce the prohibited instrumented resource effect divided by invalid action requests presented to enforcement. Execution false rejection is $\mathrm{FRR}_{\mathrm{exec}}=$ oracle-valid action requests denied, quarantined, or degraded below the specified resource success condition divided by valid action requests presented. Reliance false acceptance is $\mathrm{FAR}_{\mathrm{rel}}=$ oracle-invalid reliance cases assigned verified-for-reliance divided by invalid reliance cases presented to the independent consumer. Reliance false rejection is $\mathrm{FRR}_{\mathrm{rel}}=$ oracle-valid reliance cases assigned rejected divided by valid reliance cases presented. Under this definition, cases assigned uncertain and cases assigned escalated are not false rejections; either case fails G3 when the oracle expected verified-for-reliance.

Policy-decision accuracy is $\mathrm{PDA}=$ SUT policy dispositions matching the harness reference policy divided by policy-decision cases presented. Omission detection is $\mathrm{OD}=$ harness-injected required omissions that the validator flags as missing and the consumer assigns the oracle-expected rejected, uncertain, or escalated disposition divided by injected omissions presented. Stale-acceptance rate is $\mathrm{SAR}=$ stale or replayed items assigned verified-for-reliance divided by stale or replayed items presented.

For each revocation test, the task manifest predeclares the observation horizon $H$ as an absolute observation-end timestamp in the same clock domain as $t_{\mathrm{publish}}$ and $t_{\mathrm{reject}}$. If rejection occurs by $H$, case $i$ has $L_i=t_{\mathrm{reject}}^{(i)}-t_{\mathrm{publish}}^{(i)}$ and event indicator $\delta_i=1$. If no rejection occurs by $H$, the case is right-censored with $L_i=H-t_{\mathrm{publish}}^{(i)}$ and $\delta_i=0$. The report retains the latency lower bound and the non-rejection count $\sum_i(1-\delta_i)$ rather than dropping those cases. The artifact report records the common clock basis alongside $\mathcal{L}_{\mathrm{rev}}=\{(L_i,\delta_i)\}$.

Causal accuracy ($\mathrm{CA}$) is correctly linked oracle-required events divided by oracle-required events. Disclosure-minimization error ($\mathrm{DME}$) is non-required disclosed fields divided by all disclosed fields, so a lower value means less excess disclosure. Recovery correctness ($\mathrm{RC}$) is satisfied recovery postconditions divided by specified postconditions. Task utility ($\mathrm{TU}$) is successfully completed valid tasks divided by valid tasks presented. For each declared resource or monetary unit $k$, operational cost ($\mathrm{OC}_k$) is the total measured cost $c_k$ divided by completed valid tasks; heterogeneous units remain separate. Zero completed tasks yields N/A under the common denominator rule. Every ratio also reports its exclusion count. Inter-verifier agreement remains a future empirical requirement. A later study can define raw agreement as matching decisions divided by jointly evaluated cases, but this paper reports no value.

The score vector is
\begin{multline}
\mathbf{s}=(\mathrm{IVS},\mathrm{FAR}_{\mathrm{exec}},\mathrm{FRR}_{\mathrm{exec}},
\mathrm{FAR}_{\mathrm{rel}},\mathrm{FRR}_{\mathrm{rel}},\\
\mathrm{PDA},\mathrm{OD},\mathrm{SAR},\mathcal{L}_{\mathrm{rev}},\mathrm{CA},
\mathrm{DME},\mathrm{RC},\mathrm{TU},\{\mathrm{OC}_k\}).
\end{multline}
The score vector preserves the distinction between evidence consumption, error, timing, linkage, disclosure, recovery, utility, and cost. It also exposes cases in which local task utility is retained but cross-domain verification fails.

An artifact report contains the task and threat manifest, system configuration and component versions, repetition policy, inputs, and raw outputs. It retains the envelopes and permitted disclosure views, oracle records, validator and consumer configurations, and status histories. Challenge and recovery records, gate outcomes, and every metric numerator and denominator are also included. The report records $H$ and the shared clock basis, each revocation censor indicator and latency or lower bound, and the non-rejection count. Missing or withheld artifacts and the reason for each omission remain visible, as do cost units, endpoint outages, retries, and unresolved adjudication states.

\subsubsection{Adversarial evidence tests}

\paragraph{Integrity and temporal tests.}
The payload-tampering and signature-failure family mutates a known intact representation. Its oracle is the canonical payload, verification key, and mutation record; the validator must reject the altered or unverifiable object before policy acceptance. Selective omission removes an attestation or event required by the manifest. The validator must mark the envelope incomplete, after which the consumer returns the oracle-expected rejected, uncertain, or escalated terminal reliance disposition. Replay or stale evidence reuses an otherwise valid object outside its request identifier, nonce, audience, validity interval, or status context. Those fixed bindings form the oracle, and the verifier must reject the replay rather than treat signature validity as freshness.

\paragraph{Source, status, and policy tests.}
Equivocation or conflicting logs present incompatible signed histories to different consumers. The harness fixes the expected history and recipient-view pairs or supplies independent observations; neither an issuer-local record under test nor one of the conflicting histories is the oracle. Expected behavior requires conflict detection and the terminal reliance disposition specified by the oracle. Any quarantine or challenge action is recorded separately as an operational response. The revoked-credential family uses a credential against a harness-owned status timeline or a status attestation issued by an authority outside the evaluated system. Effective revocation and publication times are the oracle; post-revocation reliance must be rejected, while earlier reliance follows the declared historical-status policy. The compromised-producer family issues an authentic but false self-report alongside an independent counter-record. The controlled compromise and counter-record are the oracle; authentication may pass while the item must not reach verified-for-reliance. Verifier-policy mismatch supplies the same envelope to the declared policy and to a deliberately different policy or version. The manifest's policy binding is the oracle, and substitution must cause rejection or explicit escalation.

\paragraph{Disclosure and availability tests.}
The privacy-leakage family adds fields outside the audience- and purpose-specific disclosure allowlist. That allowlist and field classification are the oracle, and the disclosure view must fail even if its signatures are valid. Unavailable status or adjudication endpoints follow a controlled outage schedule and the manifest's predeclared outage rule. The verifier must return the declared terminal reliance disposition. Any quarantine, bounded degraded mode with recorded cache age and reconciliation duty, or challenge action is recorded separately as an operational response. An unavailable adjudication function cannot yield a resolved outcome and remains reported as unresolved.

These tests specify expected observations and terminal reliance dispositions under the threat assumptions in Section~\ref{sec:logical-reference-model}; they do not supply measured values or universal thresholds.

%% file: figures/fig-taas-eval-protocol.tex
\begin{figure*}[t]
\centering
\includegraphics[width=\textwidth]{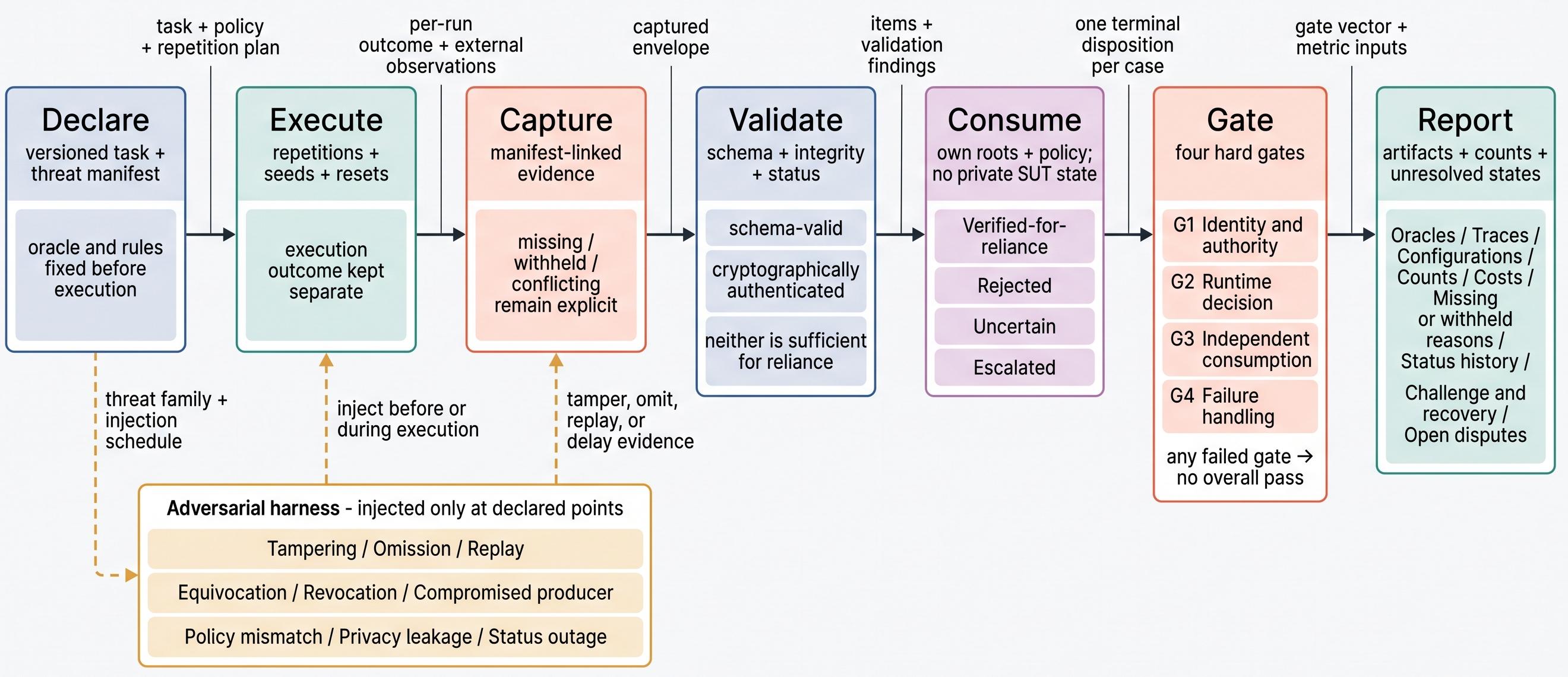}
\caption{Proposed TaaS-Eval sequence in which a harness injects adversarial conditions before or during execution and capture. Captured envelopes then pass through validation, independent consumption, hard gates, and artifact reporting.}
\label{fig:taas-eval-protocol}
\end{figure*}

%% file: tables/tab-taas-eval-protocol.tex
\begin{table*}[t]
\caption{TaaS-Eval hard gates, observable evidence, oracles, metrics, and failure rules.}
\label{tab:taas-eval-protocol}
\centering
\footnotesize
{\setlength{\tabcolsep}{2.5pt}
\renewcommand{\arraystretch}{1.12}
\begin{tabular}{@{}>{\raggedright\arraybackslash}p{0.17\linewidth}>{\raggedright\arraybackslash}p{0.31\linewidth}>{\raggedright\arraybackslash}p{0.22\linewidth}>{\raggedright\arraybackslash}p{0.24\linewidth}@{}}
\toprule
Gate/checkpoint & Observable evidence & Oracle or validator and metrics & Failure rule \\
\midrule
Task/threat manifest & Versioned scenario, boundary, SUT, task classes, policy, threat schedule, required envelope profile, repetition policy, and artifact plan. & Schema and version check; configuration digest; oracle fixed before runs; required-field coverage. & Undeclared oracle, policy, threat, or required record invalidates the task; no score is assigned. \\
G1: identity and authority & Subject credential, delegation chain, audience, action and resource scope, validity interval, status endpoint, and authority attestation. & Harness delegation matrix and external or harness-owned status timeline; correctly bound claims / required authority claims. & G1 fails if missing, revoked, audience-mismatched, or out-of-scope authority is accepted. \\
G2: runtime decision & Policy-versioned decision, request ID, enforcement and execution attestations, and instrumented resource effect for allow, deny, quarantine, degrade, or escalate outcomes. & Harness reference-policy cases, expected-action matrix, and deliberate version mismatch; $\mathrm{PDA}$, $\mathrm{FAR}_{\mathrm{exec}}$, $\mathrm{FRR}_{\mathrm{exec}}$, task utility, and operational cost. & G2 fails if no decision is linked to enforcement, policy substitution is accepted, or an oracle-invalid action proceeds. \\
G3: independent consumption & Permitted disclosure view, event and attestation IDs, causal parents, capture scope, consumer trust roots, policy, schemas, evidence access, and terminal reliance disposition. & Harness event graph, omission record, valid/invalid matrix, and expected disposition independent of producer output; $\mathrm{IVS}$, $\mathrm{FAR}_{\mathrm{rel}}$, $\mathrm{FRR}_{\mathrm{rel}}$, causal accuracy, and $\mathrm{OD}$. & G3 fails if the consumer needs private SUT state, infers completeness from missing links, or returns a non-oracle disposition. \\
G4: status, disclosure, and failure handling & Target IDs, external status history, event time, observation horizon $H$, outage schedule, disclosure allowlist, tampered or conflicting evidence, counter-records, challenge records, and recovery attestations. & External or harness-owned status timeline and expected-disposition matrix; $\mathcal{L}_{\mathrm{rev}}$, non-rejection count, $\mathrm{SAR}$, disclosure-minimization error, and recovery correctness. & G4 fails if revoked, stale, leaking, unavailable, or conflicting evidence is silently accepted, or if an unresolved conflict is reported as resolved. \\
\bottomrule
\end{tabular}
}
\end{table*}

%% file: sections/08-governance-limitations-agenda.tex
\section{Governance, Tradeoffs, Limitations, and Research Agenda}
\label{sec:governance-limitations-agenda}

\subsection{Standards and governance alignment}

The trust-evidence envelope is an integration profile, not a substitute for organizational governance, security engineering, or legal analysis. The cited instruments align with the profile at the level of function, not as interchangeable schemas. Risk frameworks and management-system instruments such as the NIST AI RMF, its generative-AI profile, ISO/IEC 42001, and ISO/IEC 23894 can identify governance, assessment, monitoring, and response functions whose decisions need evidence~\cite{nist2023airmf,nist2024genaiprofile,isoiec42001-2023,isoiec23894-2023}. Where applicable, the EU AI Act similarly raises role- and risk-dependent documentation, record, oversight, and reporting questions, but legal applicability and sufficiency remain outside the envelope~\cite{europeanunion2024aiact}.

Security catalogs and technical specifications contribute different parts of the design. OWASP LLM and Agentic Top 10 materials supply threat cases for manifests and adversarial tests~\cite{owasp2025llmtop10,owasp2026agentictop10}, while NIST SSDF supplies secure-development and response practices relevant to artifact evidence~\cite{souppaya2022ssdf}. DID, VC, and PROV specifications provide reusable identifier, issuer-verifier, credential-status, and causal-lineage semantics~\cite{w3c2022didcore,w3c2025vcdm20,w3c2013provdm}. SLSA, SPDX, and C2PA provide artifact, dependency, build-provenance, and content-provenance evidence structures~\cite{slsa2025spec,spdx2024spec301,c2pa2025spec}. None of these sources determines task delegation, action acceptability, event truth, completeness, non-equivocation, or legal effect on its own.

The mapping must be instantiated by a named policy authority for a specific reliance context. That authority selects applicable source versions, actor roles, accepted issuers, evidence scope, retention, status behavior, and review procedures. For a regulation, legal applicability and sufficiency remain questions for a competent institution. For a voluntary framework or guidance document, an organization must still decide which functions and controls apply. For a data model or technical specification, successful parsing or cryptographic verification establishes only the property tested by that mechanism. This alignment therefore provides design traceability, not a claim of clause-level compliance or certification.

\subsection{Disclosure, concentration, and surveillance}

Selective disclosure trades data minimization against evidentiary sufficiency. A tool provider may need the delegation scope and current status without receiving the user's full prompt, while an incident reviewer may need a wider causal record. The envelope addresses this tension by binding each disclosure view to an audience, reliance purpose, source manifest, included and omitted fields, transformation, evidence references, and expiry. That binding makes redaction and omission reviewable. It does not guarantee that the source record was complete, that a transformation preserved every relevant relation, or that correlation across repeated identifiers cannot reveal sensitive behavior. A relying party may reject an over-disclosing view as a privacy failure and still find a more restricted view insufficient for its decision.

A shared control plane can reduce integration work, but it can also concentrate policy, identity, status, and behavioral traces in one operator. Such concentration creates operational and governance failure modes: a service outage can block unrelated domains, a captured policy authority can authorize harmful actions under an apparently valid policy, and a high-resolution evidence store can become a surveillance system. Separating issuer, verifier, policy, evidence custody, incident coordination, and adjudication roles can limit unilateral control only when those roles have genuinely independent governance. Purpose-bound access, tenant isolation, short retention where appropriate, and audience-specific views reduce exposure but do not remove the incentive or capability to monitor agent activity. The topology-neutral reference model permits distributed allocation; it does not make decentralization an assurance property by itself.

\subsection{Interoperability, lock-in, and evidentiary failure}

Interoperability depends on semantics, not merely on serializing the same field names. Domains must agree on issuer discovery, delegation vocabulary, policy and schema versions, resource identifiers, event relations, time interpretation, status values, disclosure profiles, challenge states, and failure dispositions. A vendor-specific identifier registry, policy language, evidence locator, or status API can make an otherwise portable envelope unusable outside that vendor's service. Export functions are insufficient if another verifier cannot reproduce the relevant interpretation. Versioned issuer profiles, public conformance cases, explicit extension rules, and migration records can reduce lock-in, but the present design does not choose one transport, credential technology, or federation mechanism.

Well-formed evidence can also launder a producer's false account. A compromised runtime may sign an invented event, omit a disconfirming branch, or present different histories to different recipients. A compromised trust root can make an unauthorized issuer appear acceptable, and colluding accepted sources can construct a mutually consistent account. Signatures, hashes, schema checks, append-only records, and causal identifiers can expose alteration or inconsistency under their stated assumptions; they cannot establish truth, completeness, or source independence. A reliance policy must therefore identify when counterparty evidence, an independently governed observer, consistency checking, qualified assurance, or adjudication is required. Root compromise additionally needs an out-of-band replacement channel, distrust or revocation of descendants, preservation of the suspected interval, and reassessment of decisions made during it.

The envelope has no legal effect on its own. It can preserve a delegation record, a provider observation, a challenge, and a recovery outcome, but it cannot determine consent, breach, fault, responsibility, governing law, or remedy. Those consequences depend on applicable law, an external agreement, and institutions with authority to interpret them. Naming an adjudication path makes the dependency explicit; it does not guarantee access, independence, timeliness, or enforceability. A technical verifier must leave an unresolved dispute unresolved rather than infer legal finality from a complete-looking record.

\subsection{Operational cost and adoption}

Cross-domain evidence adds work to both the execution and governance paths. Issuers must protect keys, version schemas, bind records to events, and retain addressable evidence. Verifiers must resolve trust inputs, retrieve permitted objects, query historical status, process omissions or conflicts, and preserve their decision basis. Organizations also need incident contacts, challenge handling, recovery postconditions, and migration procedures. Selective disclosure and independent evidence may add computation, latency, storage, and human review. TaaS-Eval consequently treats operational cost as a vector of declared units rather than assuming a single monetary score, but this paper supplies no measured cost or acceptable threshold.

Adoption could be uneven if the party that bears evidence-production cost does not receive the immediate benefit. Large providers may impose proprietary profiles; smaller organizations may lack verification or adjudication capacity; counterparties may disagree about assurance scope or evidence retention. Incremental deployment can begin with a bounded action class and a small accepted issuer set, but partial adoption must not be represented as end-to-end cross-domain assurance. The profile becomes useful only when the relying party can consume the relevant records under independently administered policy and when failure states lead to predeclared dispositions rather than silent acceptance.

\subsection{Limitations}

The purposive evidence synthesis supports construction and comparison, not completeness or prevalence claims. The diagnostic and worked scenarios were produced within one analytical process, so their categories may omit reliance propositions or encode author judgment that independent coders would contest. The logical reference model identifies roles, messages, and failure states, but it does not demonstrate implementability, performance, usability, or resistance to an adaptive adversary. The evaluation protocol defines manifests, oracles, gates, metrics, and artifacts; until independent systems execute it, it yields no evidence about effectiveness or comparative cost.

Several assumptions also sit outside the envelope. Accepted trust roots, clocks, status authorities, independent evidence sources, disclosure enforcement, and adjudication procedures must be established and governed. The logical reference model records their identities and failure states but cannot create institutional independence. It does not prevent collusion among all accepted sources, repair a compromised root through that root, compel disclosure of unknown records, or ensure that a remedy is available. These are scope boundaries rather than implementation details.

\subsection{Falsifiable research milestones}

The first question is whether independently implemented domains can discover and interpret the same issuers without bilateral, vendor-specific code. A measurable milestone is a set of interoperable issuer profiles with versioned discovery metadata, verification methods, status endpoints, extension rules, and positive and negative test vectors. Independent implementations should publish per-case conformance results and preserve disagreements about unsupported versions or ambiguous ownership. The artifact is the profile, test corpus, and reproducible conformance report, not a claim that one registry is universally authoritative.

A second question concerns task-scoped delegation semantics. Can a verifier determine the permitted subject, audience, task, action, resource, constraints, time, attenuation chain, and revocation consequence from records issued by different domains? The milestone is a versioned delegation vocabulary and test suite containing valid chains, over-broad and attenuated delegations, audience substitution, replay, expiry, revocation, and conflicting interpretations. Expected dispositions and oracle inputs should be fixed before implementations are evaluated. Cases that require institutional interpretation must receive the escalated terminal reliance disposition and pass to adjudication rather than being scored as technically resolved.

Provenance work must test equivocation rather than equate append-only storage with consistency. The research question is whether a consumer can detect omitted, replayed, or split-view records when one producer controls its own history. An equivocation-resistant provenance test corpus should include recipient-specific conflicting views, counterparty receipts, independent observations, causal-parent mutations, and issuer collusion assumptions. Each case should name what the consumer can observe and require the expected rejected, uncertain, or escalated terminal reliance disposition. Any quarantine or challenge action should be recorded separately as an operational response. A useful result would report detection and non-detection by threat class, with the producer's disputed history excluded as the sole oracle.

Selective disclosure requires its own verification artifacts. Can a consumer verify audience and purpose binding, linkage to the committed source, declared omissions, transformation, and expiry without receiving non-required content? A milestone is a disclosure-policy suite with permitted and prohibited fields, over-disclosure and under-disclosure cases, transformed views, unavailable evidence, and correlation-risk annotations. Implementations should report field-level disclosure outcomes and reliance dispositions. This would test both privacy leakage and evidentiary sufficiency without assuming that cryptographic correspondence establishes source completeness.

Evaluation reliability remains empirical. Independent evaluator agreement studies should give multiple verifiers the same task manifest, envelope views, accepted trust inputs, policy version, and oracle-access rules. They should report raw agreement, disposition-specific confusion counts, and coded reasons for disagreement for each terminal reliance disposition: verified-for-reliance, rejected, uncertain, and escalated. The study protocol should distinguish differences caused by implementation defects, policy interpretation, unavailable evidence, and adjudicative judgment. No agreement threshold is proposed here; the measurable artifact is a preregistered decision set with per-case explanations and repeatable analysis.

Recovery must be tested against observable postconditions rather than a producer's declaration of success. Can a relying party establish that authority was revoked, affected actions were bounded, required state was repaired, status was republished, and follow-up obligations were completed? Recovery postcondition tests should pair instrumented resource effects and external status timelines with false self-reports, partial cleanup, delayed publication, endpoint outage, and conflicting counter-records. Reports should retain each postcondition, its evidence source, verification time, and unresolved status. A recovery attestation passes only the checks defined for that case; it does not erase the original action or decide the adequacy of a remedy.

The final milestone is a family of cross-domain conformance profiles that combines these artifacts for bounded workflow classes. Each profile should name required and optional attestations, issuer independence assumptions, disclosure views, status and challenge behavior, failure dispositions, and applicable framework mappings. Shared manifests and test vectors should be run across centralized, federated, and protocol-mediated deployments to determine whether topology changes the observable reliance semantics. Conformance reports must expose omitted tests, unsupported extensions, unavailable institutions, and unresolved disputes. Such profiles would provide evidence about interoperability within the tested workflow classes and describe conditions relevant to adoption while preserving the distinction between technical conformance, policy acceptance, and legal effect.

%% file: sections/09-conclusion.tex
\section{Conclusion}
\label{sec:conclusion}

Cross-domain agentic workflows create a reliance problem that local safeguards cannot settle for another administrative authority. A runtime can authenticate users, constrain tools, enforce policy, and retain logs inside its own domain, yet a provider or counterparty still needs a defensible basis for deciding whether a particular action was delegated, remained in scope, produced the asserted effect, and retained valid status. TaaS integrates those action-specific relations around a named relying party and administrative trust boundary; it does not replace identity, authorization, provenance, assurance, governance, or incident-response mechanisms.

The paper supplies a bounded design blueprint. The structural-externalization diagnostic identifies propositions that combine multi-domain dependence, independent reliance, and adverse-condition continuity. The trust-evidence envelope binds an immutable manifest to issuer-attributed authority, decision, execution, provenance, status, disclosure, challenge, and recovery records. Omissions and disagreements remain explicit. The logical reference model assigns those functions to explicit roles and trust domains without requiring a central platform. Three analytical scenarios and the TaaS-Eval protocol proposal then specify how future implementations can expose independent consumption, tampering, replay, revocation, equivocation, privacy, recovery, and cost outcomes.

Future work should implement and conformance-test the profile, measure evaluator agreement, and assess it with stakeholders and relevant institutions. These studies should determine what evidence can be established about event truth, completeness, policy acceptance, responsibility, and legal effect when trust roots, evidence sources, status services, or adjudication institutions fail or collude. Until then, the paper makes design claims rather than operational claims.